\documentstyle[aps,twocolumn,pre,epsfig]{revtex}
\begin{document}
\draft

\title{A pseudo-spectral approach to inverse problems in
interface dynamics}
\author{Achille Giacometti$^{(1)}$, and Maurice Rossi$^{(2)}$}

\vskip 2.0cm
\address{$^{(1)}$
INFM Unit\'a di Venezia, Dipartimento
di Chimia Fisica, Universit\'a di Venezia,\\
Calle Larga Santa Marta DD 2137, I-30123 Venezia-Italy}
\address{$^{(2)}$ Laboratoire de Mod\'elisation en M\'ecanique,
Universit\'e de Paris VI, 4 Place Jussieu, F-75252 Paris Cedex 05, France}

\date{\today}

\maketitle

\begin{abstract}
An improved scheme for computing coupling parameters of the
Kardar-Parisi-Zhang equation from a collection of successive
interface profiles, is presented. The approach hinges on
a spectral representation of this equation. An appropriate
discretization based on a Fourier representation, is discussed
as a by-product of the above scheme.
Our  method is first tested on profiles generated by a one-dimensional
Kardar-Parisi-Zhang equation where it is shown to reproduce the input
parameters very accurately. When applied to microscopic models of growth,
it provides the values of the coupling parameters associated with the
corresponding continuum equations. This technique favorably compares 
with previous methods based on real space schemes.
\end{abstract}

\section{Introduction}

In most complex systems, it is often difficult to directly  relate microscopic
interactions  to  the dynamics of  coarse-grained~(mesoscale or
large scale)  spatial structures.   In this context, nonlinear inverse
methods~\cite{Weigend94}   which infer equations governing a system from
experimental observations of its successive time  evolution,  may prove to  be
more efficient then direct methods.  When  only experimental data
 coupled with the
hypothesis of an underlying determinism are used, the identification is
purely non-parametric. Such  methods are extensively exploited 
for instance in biological and
economical systems where predictions are typically relying  not on basic
mechanisms, but  directly extrapolated from time series using 
various procedures 
(neural networks, nearest-neighbors algorithms, etc) ~\cite{Weigend94}. 
On the other hand if a
parametrized phenomenological equation, usually derived from a 
combination of general
symmetry considerations and  heuristic physical arguments, is assumed from the
outset, the inverse approach   consists in   finding an optimal 
set of parameters.  In
this work, we   focus on this latter situation in the framework of
interface growth dynamics.

The   phenomenon of  interface growth  covers many technological
applications ranging from epitaxial deposition to bacterial growth and
fluid motion in
porous media~\cite{Krug97,Barabasi95,Halpin95,Marsili96,Meakin93}. 
It is known that the large scale dynamics of such
systems
may   be conveniently address in terms of  continuum stochastic differential
equations. The Kardar-Parisi-Zhang~(KPZ) equation~\cite{Kardar86} constitutes a
paradigmatic universality class which is believed to aggregate a large portion
of observed  interface dynamics. This  Langevin type equation~\cite{Risken89}
possesses  a non-linear term which accounts for the interface local normal
growth
absent in  its linear  counterpart the Edward-Wilkinson
equation~(EW)~\cite{Edward82}. Despite a major effort both from the
numerical and
analytical point of view, a complete characterization of the KPZ properties
is nevertheless still
lacking.  From the numerical viewpoint, finite-difference schemes have been
widely exploited  to approximate the continuum   KPZ  dynamics. 
Unfortunately, naive
discretizations of the  spatial derivatives may result into   behaviors
inconsistent with known properties of the continuum equation.  
For instance, the
usual symmetric second-order finite difference scheme for the non-linear term
violates an important symmetry of the continuum one-dimensional
theory~\cite{Lam98_1}.  An appropriate modification in the framework of
finite difference approximations may  heal  this
drawback~\cite{Lam98_1,Lam98_2}, but it is nonetheless
unable to properly display   coarse-grained properties
of the original continuum equation~\cite{Giacometti00}.
Specifically it does not  preserve
the  correct functional form of the coarse-grained  
equilibrium distribution, a basic feature that one expects from
the renormalization group point of view  (a similar feature
was already observed in Ref~\cite{Giacometti95}).

In the present work, we follow a different route by introducing a numerical
approach which preserves  more  features of the original 
continuum KPZ equation. From the
outset, our method is based  on a spectral rather than a 
finite difference scheme.
Spectral methods are widely used  in fluid mechanics~\cite{Canuto88} and their
accuracy and reliability compared  to those of finite difference
schemes have been tested, over the years, on deterministic equations. 
Either in the
context of fluctuating interfaces, it has already 
been used ~\cite{Zaleski89,Hayot93} 
to analyze the evolution of a deterministic  equation, namely   
the Kuramoto-Shivashinsky equation.
However, it has not been directly implemented on
the stochastic partial differential equation \cite{Toral00}. 
In this
paper, we first extend spectral methods to stochastic equations, using the KPZ
equation as a testbench, then devise a reconstruction procedure
for the inverse problem for the KPZ, and finally show that it
is unaffected by the deficiencies of real space approximations.

The first   reconstruction of the stochastic  KPZ dynamics from experimental
surface  has been performed by  Lam and Sander~\cite{Lam93}.  These authors
used a  standard finite-difference scheme  to approximate the dynamics and
based their reconstruction on  a  least-squares  algorithm.
Experimental dynamics was obtained   by simulating various  
microscopic models.
This strategy, however, seems  to be limited  for two reasons.
First  it is based on a problematic  numerical approximations of the KPZ
equation. Second  the least-squares algorithm, originally tailored for
deterministic equations was directly transpose to stochastic or Langevin type
equations~\cite{Lam93}. In fact, while 
the performance of this method has been tested in
the presence of  measurement noise - noise which   affects   experimental
observations  but does not change   deterministic trajectories -  
in the presence of dynamical noise (e.g. Langevin dynamics) 
this scheme  can be
far more unsuccessful. Moreover, the
least-squares algorithm of Ref.~\cite{Lam93} is very  much dependent on the
assumption of  small sampling time  of observations a  fact that should be
checked {\it a posteriori} (see Ref. [12] for more details).

In a previous paper~\cite{Giacometti00},  an alternative reconstruction
algorithm  which  does not suffer of  the  aforementioned  problems,  was
introduced.  The basic strategy amounts to apply the least-squares
procedure at the
level of correlation functions rather than directly to 
the interface stochastic
variables. This method, albeit successful, could not  correctly  
account for the
surface coarse grained properties  because of an intrinsic deficiency of 
spatial discretizations. 
Motivated by this feature, we shall remove this drawback by introducing
a spectral representation of the Langevin equation
which is not plagued by discretization problems.
Then we shall proceed to reformulate and test the approach of Ref~[12]
for the inverse KPZ problem,
in spectral space.

The paper is roughly divided in two parts and organized as follows. 
The first part  concerns the KPZ
equation itself and its approximations. 
In Section~\ref{sec:KPZ}, the continuum KPZ equation in ($1+1$)-dimension
is rapidly recalled along with some previous spatial
discretizations used to
numerically mimic it. In the same Section, a novel
discretization
in Fourier space  is then proposed and shown to 
improve over the real space approach from a purely theoretical viewpoint.
The deterministic evolution equation for correlations functions 
in spectral space, are then derived in  Section~\ref{sec:correlations}.  
Section~\ref{KPZnumeric} is  devoted  to the numerical
procedure and, in particular, to the treatment, {\it via } a
pseudo-spectral method, of  the nonlinear term of the KPZ  equation. 
Numerical results are then provided  to compare   
the performance of the various discretizations to the
corresponding   analytical results obtained for the continuum KPZ equation.  
The second part of the paper  is    devoted  to the 
inverse method for KPZ dynamics, where  we discuss how one can 
reconstruct the dynamics from the knowledge of interface profiles.
This technique, based on   equations derived in
Section~\ref{sec:correlations},   is  described in Section~\ref{sec:ident}. 
It is then  tested  in Section~\ref{KPZpseudoreconstr} against 
data produced by synthetic interface profiles generated 
by a ($1+1$)-dimensional
KPZ equation with known coupling parameters. The test is performed even
in the presence of coarse-graining and provides the renormalization 
properties of the KPZ coupling parameters.
Finally, in Section~\ref{sec:Micro}, the reconstruction method 
is applied  to various  microscopic models  at various coarse-grained scales.

\section{Spectral  discretization of the KPZ equation }
\label{sec:KPZ}

We consider a one-dimensional surface profile of width $L$. The surface
can be either experimentally or numerically generated, and we assume it
to be periodic with period $L$. At a mesoscopic scale, it can be
described within a continuum (hydrodynamic) approximation by a variable
$h(x,t)$ (with $0 \le x \le L$) which satisfies a Langevin equation 
\cite{Krug97}. A paradigmatic example is given by the Kardar-Parisi-Zhang
(KPZ) equation \cite{Kardar86}:

\begin{eqnarray}
\partial_t h  &=& c  + \nu  \partial_x^2 h
+\frac{\lambda }{2} (\partial_x h )^2 + \eta, 
\label{exp:kpzcon}
\end{eqnarray}
where  $ \eta (x,t) $ is a noise with zero average and $\delta$-correlated
both in time and space as:
\begin{eqnarray}
\left \langle \eta (x,t) \eta (x',t') \right \rangle &=& 2 D
\delta(x-x')~\delta(t-t').
\label{exp:deltaxt}
\end{eqnarray}
In Eq.~(\ref{exp:kpzcon}), symbol  $\left \langle \ldots  \right\rangle $ means
that   an ensemble average over the noise is performed and
$c$, $\nu$, $\lambda$, and $D$ are coupling parameters.
In writing Eqs.(\ref{exp:kpzcon}) and (\ref{exp:deltaxt}), an appropriate
regularization is always tacitly assumed.   This amounts to define  
a minimal length scale   $a$  for  $h(x,t)$
 and to assume that the surface variable $h(x,t)$  satisfies
a KPZ regularized equation with
a given spatial cut-off depending on scale $a$. For instance, this
can be done by  considering a discretization version of Eqns.
(\ref{exp:kpzcon}) and (\ref{exp:deltaxt}) with $a$ as a lattice constant :

\begin{eqnarray}
\frac{d h_i }{dt}&=& c+ \frac{\nu}{a^2}
F_i^{\nu}[h] + \frac{\lambda}{2 a^2} \bar{F}_i^{\lambda}[h] +
\sqrt{\frac{D}{a}}  \theta_i,
\label{exp:kpzdiscr1}
\end{eqnarray}
or alternatively

\begin{eqnarray}
\frac{d h_i }{dt} &=& c+ \frac{\nu}{ a^2}  F_i^{\nu}[h] +
\frac{\lambda}{2 a^2} F_i^{\lambda}[h]  +  \sqrt{\frac{D}{a}}  \theta_i,
\label{exp:kpzdiscr2}
\end{eqnarray}
 where $h_j$ stands for the value   $h(x_j,t)$  of the periodic smoothed
surface at $x_j=ja$, $j=1,\ldots,N$ where  $N=L/a$ and  the quantities
$\theta_i(t)$ are uncorrelated white noise functions
\begin{eqnarray}
\left \langle \theta_i (t) \theta_j (t') \right \rangle &=&
2 \delta_{ij}  \delta(t-t')~.
\label{exp:deltait}
\end{eqnarray}
In Eqs.~(\ref{exp:kpzdiscr1}) and (\ref{exp:kpzdiscr2}),  one  usually
approximate the discretized Laplacian with the second-order finite difference 
term
\begin{eqnarray}
F_i^{\nu}[h]&=& h_{i+1}+h_{i-1}-2 h_i,
\label{exp:laplacian}
\end{eqnarray}
and  the non-linear term with
\begin{eqnarray}
\bar{F}_i^{\lambda}[h]&=& \frac{1}{4} \bigl[ h_{i+1}-h_{i-1} \bigr]^2,
\label{exp:standard}
\end{eqnarray}
or
\begin{eqnarray}
F_i^{\lambda}[h]&=&\frac{1}{3}  \bigl[(h_{i+1}-h_{i})^2+(h_{i}-h_{i+1})^2 \\ \nonumber
&+&
(h_{i+1}-h_i)(h_i-h_{i-1})\bigr].
\label{exp:lam-shin}
\end{eqnarray}
As  previously discussed ~\cite{Lam98_1,Lam98_2,Giacometti00},
only choice given by Eq.~(\ref{exp:lam-shin}) 
guarantees that at least some properties of the correct equilibrium
distribution are retrieved. 
{\em Both} representations Eqs.~(\ref{exp:standard}) and
~(\ref{exp:lam-shin})  
are however problematic at the coarse-grained
level \cite{Giacometti00}.

Here we show how one can achieve an alternative and {\it always} correct
discretization of the continuum KPZ equation by a procedure that
directly applies in momentum space.

We first expand the periodic continuum field $h(x,t)$   in Fourier modes as
\begin{eqnarray}
h(x,t) &=& \frac{1}{L} \sum_{n=-\infty}^{+\infty} \widehat{h}_{q_{n}}(t)~
e^{i q_n x},
\label{four1}
\end{eqnarray}
where the  Fourier component
\begin{eqnarray}
\widehat{h}_{q_{n}}(t) &=&\int_{-L/2}^{L/2}~dx~h(x,t)~e^{-i q_n x},
\label{four2}
\end{eqnarray}
is associated with   wavenumber $q_n =2\pi n/L$. Since  $h(x,t)$ is
real, this imposes  $\widehat{h}_{q_{n}}^*=\widehat{h}_{-q_{n}}$ or 
alternatively, if
$\widehat{h}_{q_{n}} = \widehat{\alpha}_{q_{n}} +i
\widehat{\beta}_{q_{n}}$  is separated in real and imaginary part,
$\widehat{\alpha}_{-q_n}=\widehat{\alpha}_{q_n}$,
$\widehat{\beta}_{-q_n}=-\widehat{\beta}_{q_n}$.   A similar expansion for  the
noise term $\eta(x,t)$ leads to   Fourier  components  $\widehat{\eta}_{q_n}$
with the following correlations~: 
\begin{eqnarray}
\left \langle \widehat{\eta}_{q_n}(t) \widehat{\eta}_{q_m}(t') \right
\rangle &=&2 D L \delta_{n,-m} \delta(t-t').
\label{corr1}
\end{eqnarray}
Again by decomposing   Fourier modes $\widehat{\eta}_{q_n} \equiv
\widehat{\xi}_{q_{n}} + i \widehat{\zeta}_{q_{n}} $   in their real and
imaginary parts,  one obtains, in addition to    relations
$\widehat{\xi}_{q_n}= \widehat{\xi}_{q_{-n}}$,
$\widehat{\zeta}_{q_n}=-\widehat{\zeta}_{q_{-n}}$,    conditions on     the
correlations
 \begin{eqnarray} 
\left \langle \widehat{\xi}_{q_n}(t) \widehat{\xi}_{q_m}(t')
\right \rangle &=&   D L \delta_{n,m} \delta(t-t') ,
\label{corr2}
\end{eqnarray}

\begin{eqnarray}
\left \langle \widehat{\zeta}_{q_n}(t) \widehat{\zeta}_{q_m}(t') \right
\rangle &=&   D L \delta_{n,m} \delta(t-t'),
\label{corr3}
\end{eqnarray}

\begin{eqnarray}
\left \langle \widehat{\xi}_{q_n}(t) \widehat{\zeta}_{q_m}(t') \right
\rangle &=&0,
\label{corr4}
\end{eqnarray}
for $n>0$,$m>0$. For the Fourier component $n=0$, one gets  
$\widehat{\zeta}_{q_{0}}=0$ and

\begin{eqnarray}
\left \langle \widehat{\xi}_{q_{0}}(t) \widehat{\xi}_{q_{0}}  (t') \right
\rangle &=& 2 D L \delta(t-t')
\label{correl0}
\end{eqnarray}
Using Eq.~(\ref{four1}) in Eq.~(\ref{exp:kpzcon}),
an infinite system of  coupled   Langevin
equations is obtained~:

\begin{eqnarray}
\frac{d \widehat{h}_{q_{n}}(t)}{dt} &=& c L \delta_{n,0}-\nu q_n^2 ~
\widehat{h}_{q_{n}}(t) \\ \nonumber 
&-&
\frac{\lambda}{2L} \sum_{m, m' =-\infty}^{\infty} q_m q_{m'}
\widehat{h}_{q_{m}}(t)
\widehat{h}_{q_{m'}}(t) \delta_{n,m+m'} + \widehat{\eta}_{q_{n}}(t).
\label{exp:kpzfour}
\end{eqnarray}
The spectral approximation now amounts to project the above infinite system
on the space of
periodic functions of
period $L$ with  a finite number of  Fourier modes $\widehat{h}_{q_{n}}$
($| q_{n}
| \le  q_{N/2}$).  All  equations retain their
original form with
the {\it proviso} that infinite sums $\sum_{n=-\infty }^{\infty }$ are now
replaced by
finite ones $\sum_{n=-N/2 }^{N/2 }$.
 This  procedure thus  assumes
that $\widehat{h}_{q_{n}}=0$ for any $n > N/2$, and the original continuum
equation is then reduced  to a
set of $N+1$ real Langevin equations.  The first one

\begin{eqnarray}
\frac{d \widehat{h}_{q_{0}}}{dt} &=& cL + \frac{\lambda}{  L}
\sum_{n=1}^{N/2}  q_n^2 \Bigl[
\widehat{\alpha}_{q_{n}}^2 +\widehat{\beta}_{q_{n}}^2 \Bigr] +
\widehat{\xi}_{q_{0}},
\label{exp:kpzfourf0}
\end{eqnarray}
governs  the temporal evolution  of the zeroth mode
$\widehat{h}_{q_{0}}$ and depends  on $\widehat{h}_{q_{n}}$
for $ 0 < n \le N/2 $. The other $ N $ real equations are  independent of
$\widehat{h}_{q_{0}}$ and describe the time evolution of
$\widehat{\alpha}_{q_1}$,
$\widehat{\beta}_{q_1}$,$\ldots$,
$\widehat{\alpha}_{q_{N/2}}$,$\widehat{\beta}_{q_{N/2}}$~:

\begin{eqnarray}
\frac{d \widehat{\alpha}_{q_n}}{dt} &=& F_{q_n}
\Bigl[\widehat{\alpha},\widehat{\beta}\Bigr]+ \widehat{\xi}_{q_{n}},
\label{exp:kpzfouralpha}
\end{eqnarray}

\begin{eqnarray}
\frac{d \widehat{\beta}_{q_n}}{dt} &=& G_{q_n}
\Bigl[\widehat{\alpha},\widehat{\beta}\Bigr]+ \widehat{\zeta}_{q_{n}},
\label{exp:kpzfourbeta}
\end{eqnarray}
where

\begin{eqnarray}
F_{q_n}\Bigl[\widehat{\alpha},\widehat{\beta}\Bigr] &=&
 -\nu q_n^2  \widehat{\alpha}_{q_n}  \\ \nonumber
&-&
\frac{\lambda}{2L} \sum_{m,m'=-N/2}^{N/2} q_m q_{m'}
\Bigl[ \widehat{\alpha}_{q_m}\widehat{\alpha}_{q_{m'}}-
\widehat{\beta}_{q_m} \widehat{\beta}_{q_{m'}} \Bigl] \delta_{n,m+m'},
\label{Fk}
\end{eqnarray}

\begin{eqnarray}
G_{q_n}\Bigl[\widehat{\alpha},\widehat{\beta}\Bigr]  &=&
-\nu q_n^2  \widehat{\beta}_{q_n} \\ \nonumber
&-&
\frac{\lambda}{2L} \sum_{m,m'=-N/2}^{N/2} q_m q_{m'}
\Bigl[ \widehat{\alpha}_{q_m}\widehat{\beta}_{q_{m'}}+
\widehat{\beta}_{q_m} \widehat{\alpha}_{q_{m'}} \Bigl] \delta_{n,m+m'}.
\label{Gk}
\end{eqnarray}
The philosophy underlying  this regularization is
akin a renormalization group scheme in momentum space, where high
wavevectors are integrated  out above  a cut-off  $q_{N/2}$ in momentum space. 
This  approximation  is an alternative - albeit not equivalent - method to
discretize, in real space, the width $L$ into $ N+1$ independent points 
separated by a
distance $a=L/N$. For  the sake of simplicity, 
the number $N$ is hereafter  assumed  
to be a power of two.

For  the  continuum  linear counterpart - the EW equation -  no approximations
are involved  in
the framework of this spectral method, as  equations for   modes  
$q_n  > q_{N/2}$ are simply
discarded.  On the other hand, this approach is shown
to be far more useful 
than the real space Eqs.~(\ref{exp:kpzdiscr1}) and (\ref{exp:kpzdiscr2}),
in the non-linear KPZ case, since it preserves, unlike
both approximations given in Eqs.~(\ref{exp:kpzdiscr1}) 
and (\ref{exp:kpzdiscr2}), some 
basic properties of the original KPZ continuum equation as shown below.
Indeed let us recall that, apart from a normalization
factor, the steady
state distribution of   continuum KPZ is given by

\begin{eqnarray}
P_s [h]  &\sim&  \exp \biggl[- \frac {\nu}{2 D}  \int_{-L/2}^{L/2}~dx
\bigl(\partial_x h \bigr)^2 \biggr].
\label{exp:kpzsteady}
\end{eqnarray}
 The distribution of  modes $|q_n| \le q_{N/2}$   in momentum space thus  reads~:
\begin{eqnarray}
P_s [\widehat{h}_{q_{1}},..,\widehat{h}_{q_{N/2}}]  &\sim&    
\exp \biggl[- \frac{\nu}{2L D}
\sum_{n=-N/2 }^{N/2} q_{n}^2 |\widehat{h}_{q_{n}} |^2 \biggr].
\label{exp:kpzsteadyfou}
\end{eqnarray}
The zeroth mode $\widehat{h}_{q_{0}}$ does not  contribute
to~(\ref{exp:kpzsteadyfou})
since the average value of $h(x,t)$ does not appear
in~(\ref{exp:kpzsteady}), meaning that the
surface always  grows and  its average   never settles to a steady value
unlike surface  gradients.
We note that both the original distribution Eq.~(\ref{exp:kpzsteady})
and its spectral approximation Eq.~(\ref{exp:kpzsteadyfou})
are {\em independent} of $\lambda$.
As already remarked in Ref. ~\cite{Lam98_1}, this  property is
not satisfied by  the  finite difference
approximation Eq.~(\ref{exp:kpzdiscr1}). This inconvenient point is partly
solved by the modification given in Eq. ~(\ref{exp:kpzdiscr2}) 
which leads to  the  correct steady state distribution
\begin{eqnarray}
P_s [h_1,\ldots,h_N]   &\sim&   \exp \biggl[- \frac {\nu}{2 D a} 
\sum_{j=1 }^{N }({h_{j+1}-h_{j}})^2 \biggr].
\label{exp:steadydiscr2}
\end{eqnarray}
However it has been  shown in Ref.~\cite{Giacometti00}, that even
the approximation given in Eq.~(\ref{exp:kpzdiscr2}), 
fails in the presence of coarse-graining. This means that 
if fluctuating interfaces, obtained by the numerical generation
of a real space discretized KPZ equation at scale $a$,  
are   smoothed out up to a scale
$a_s>a$,  these  coarse-grained surfaces cannot be  described by  a
renormalized KPZ equation
at  $a_s$. The origin of this problem can be traced back to the
form of the steady state
distribution~(\ref{exp:steadydiscr2}) within a real
space scheme. On the contrary, our
spectral discretization has been devised in such a way to avoid this
drawback (see below), and can then act as a safe starting point for the
reconstruction procedure.

On remembering that there are only $N$ independent modes, the
Fokker-Planck equation governing the evolution of the probability
distribution $P[\widehat{\alpha},\widehat{\beta},t]$  associated with the
Langevin equations~ (\ref{exp:kpzfouralpha})-(\ref{exp:kpzfourbeta}), 
reads~\cite{Risken89}:
\begin{eqnarray}
\frac{\partial   P}{\partial t} &=& \sum_{n=1}^{N/2}
[- \frac{\partial (F_{q_n}P)}
{\partial \widehat{\alpha}_{q_n}} -
 \frac{\partial (G_{q_n} P)}{\partial \widehat{\beta}_{q_n}} +  \frac{DL}{2}
(\frac{\partial ^2
P }{\partial \widehat{\alpha}_{q_n}^2}   +  \frac{\partial ^2 P}{\partial
\widehat{\beta}_{q_n}^2} ) ].
\label{Pfou}
\end{eqnarray}
One may then check that the steady solution~(\ref{exp:kpzsteadyfou})
satisfies~ Eq. (\ref{Pfou}).  It is known~\cite{Risken89} that, when such a
steady   probability   exists,   all time dependent distributions 
asymptotically converge  towards
it. The discretized KPZ  thus  preserves the particular symmetry 
of the  continuum KPZ equation. 
A further advantage of this discretization, is that surfaces coarse-grained
at length scales $a_s > a$ can be simply obtained by cutting out
modes with wavenumber larger than $q_{{N_s}/2}$ ($N_s=Na/a_s$).
From Eq. ~(\ref{Pfou}), it is straightforward to get  the steady
probability of  these coarse-grained surfaces, that is a steady probability of
the same form as
equation~(\ref{exp:kpzsteadyfou})  with $N=N_s$ and the same
$\frac{D}{\nu}$ ($L=N_s a_s$). The  coupled
Langevin equations~(\ref{exp:kpzfouralpha}) and (\ref{exp:kpzfourbeta}),
ensure that, under coarse-graining, the exact steady-state is
recovered.

\section{Evolution equations for correlation functions. }
\label{sec:correlations}
Next we address the time dependent  distribution
$P[\widehat{\alpha},\widehat{\beta},t]$ appearing in equation~(\ref{Pfou}).
We first   exhibit
an evolution equation for the ensemble average
$\left \langle \widehat{\alpha}_{q_n}^2\right \rangle (t) $,  and
$ \left \langle \widehat{\beta}_{q_n}^2 \right \rangle (t) $
\begin{eqnarray}
\left \langle \widehat{\alpha}_{q_n}^2 \right \rangle (t)&=&
\int {\cal D}\widehat{\alpha}{\cal D}\widehat{\beta}~
\widehat{\alpha}_{q_n}^2~ P[\widehat{\alpha},\widehat{\beta},t],
\end{eqnarray}
\begin{eqnarray}
\left \langle \widehat{\beta}_{q_n}^2 \right \rangle (t)&=&
\int {\cal D}\widehat{\alpha}{\cal D}\widehat{\beta}~
\widehat{\beta}_{q_n}^2~ P[\widehat{\alpha},\widehat{\beta},t],
\end{eqnarray}
where
\begin{eqnarray}
{\cal D}\widehat{\alpha}{\cal D}\widehat{\beta} &=&\prod_{j=1}^{N/2}
~d\widehat{\alpha}_{q_j}~d\widehat{\beta}_{q_j}.
\end{eqnarray}
Assume that  the  probability density
$P[\widehat{\alpha},\widehat{\beta},t]$   goes
exponentially  to zero for  $\widehat{\alpha}$, $\widehat{\beta}$ going to $\infty$. If  the Fokker Planck
equation~(\ref{Pfou}) is
multiplied  by $\widehat{\alpha}^2_{q_n} $ and then integrated over all
variables, one obtains,
after an integration by  parts~:

\begin{eqnarray}
{\frac { d \left \langle \widehat{\alpha}_{q_n}^2  \right \rangle (t)} {dt} }
&=& 2 \int ~ {\cal D}\widehat{\alpha}{\cal D}\widehat{\beta}
~ \widehat{\alpha}_{q_n} F_{q_n}[\widehat{\alpha},\widehat{\beta}]
P[\widehat{\alpha},\widehat{\beta},t] +DL,
\label{ensaverage1}
\end{eqnarray}
and similarly for $\widehat{\beta}_{q_n}^2 (t)$,

\begin{eqnarray}
{\frac { d \left \langle \widehat{\beta}_{q_n}^2 \right \rangle (t) } {dt} }
&=& 2 \int ~ {\cal D}\widehat{\alpha}{\cal D}\widehat{\beta}
~ \widehat{\beta
}_{q_n} G_{q_n}[\widehat{\alpha},\widehat{\beta}]
P[\widehat{\alpha},\widehat{\beta},t] +DL.
\label{ensaverage2}
\end{eqnarray}
From the expressions (\ref{Fk})-(\ref{Gk}), we then get for $n > 0$

\begin{eqnarray}
\frac{d}{d~t} \left \langle |\widehat{h}_{q_n}|^2 \right \rangle&=&
-2 \nu q_n^2 \left \langle |\widehat{h}_{q_n}|^2  \right \rangle -
 \lambda {\cal V}_{q_n}+ 2DL,
\label{ensaverage3}
\end{eqnarray}
where
\begin{eqnarray}
{\cal V}_{q_n} &=& \frac{1}{L}  \sum_{m,~m'=-N/2}^{N/2} q_m~q_{m'}
{\cal R} \Bigl[ \left \langle  \widehat{h}_{-q_n} \widehat{h}_{q_m}
\widehat{h}_{q_{m'}}   \right \rangle \Bigr] \delta_{n,m+m'},
\end{eqnarray}
and  ${\cal R}$ indicates that only the real part is considered. 
Equation (\ref{ensaverage3}) 
implies that

\begin{eqnarray}
\frac{d}{dt} g_2(t)&=&-2 \nu g_4(t)- \lambda  K(t)+2D Q_{2},
\label{ensaverage4first}
\end{eqnarray}
where
\begin{eqnarray}
g_{2p}(t)&\equiv& \frac{1}{L} \sum_{n=-N/2}^{N/2} q_n^{2p} \left \langle
|\widehat{h}_{q_n}(t)|^2 \right \rangle,
\end{eqnarray}
for $p=1,2,\ldots$ and
\begin{eqnarray}
K(t)&=& \frac{1}{L^2} \sum_{n, m, m'=-N/2}^{N/2} q_n^2 q_m q_{m'} \\ \nonumber
& &
{\cal R} \Bigl[ \left \langle  \widehat{h}_{-q_n}(t) \widehat{h}_{q_m} (t)
\widehat{h}_{q_{m'}}(t) \right \rangle \Bigr] \delta_{n,m+m'},
\end{eqnarray}
\begin{eqnarray}
Q_{2} &=&   \sum_{n=-N/2}^{N/2} q_n^2 = \frac{4}{3} \pi^2 
\frac{N(N+1)(2N+1)}{L^2}.
\end{eqnarray}
In Appendix~\ref{sec:appA} is shown that the term $K(t)$  
in Eq.~(\ref{ensaverage4first}) actually   
vanishes identically. Hence one is left with 

\begin{eqnarray}
\frac{d}{dt} g_2(t)&=&-2 \nu g_4(t) +2D Q_{2}.
\label{ensaverage4}
\end{eqnarray}
Therefore only $\nu$ and $D$ explicitly appear in the above equation.
The absence of the parameter $\lambda$ in Eq. (\ref{ensaverage4})
is reminiscent of an analogous  phenomenon appearing in the steady probability
distribution~(\ref{exp:kpzsteadyfou}) in $1+1$ dimensional case.
It is worth stressing that, although the $\lambda$ term does not
explicitly appear in Eq.~(\ref{ensaverage4}), it is nevertheless implicitly
present through the evolution of $g_4(t)$.

Another equation can be obtained by averaging Eq.~(\ref{exp:kpzfourf0})
\begin{eqnarray}
\frac{d}{dt} g_0(t)&=& cL +\frac{\lambda}{2} g_2(t).
\label{ensaverage6}
\end{eqnarray}
where
\begin{eqnarray}
g_0(t)&\equiv & \left \langle \widehat{h}_{q_0}(t) \right \rangle,
\end{eqnarray}
This latter relation  does  not  explicitly involve  either the noise or
the diffusion term.

\section{The Pseudo-spectral method for the KPZ equation}
\label{KPZnumeric}

\subsection{Numerical procedures}

In order to   compare rough surfaces numerically generated by various  spatial
discretizations of the  KPZ equation,  we introduce   a one  step Euler
Scheme in time for
the temporal discretization.    This simple algorithm   is used  since, for
stochastic equations, it is known that a naive application
of a two-steps method 
can result in less computational  efficiency ~\cite{Mannella89}.
In order to speed up the evolution
of equations~(\ref{exp:kpzfouralpha}) and (\ref{exp:kpzfourbeta}),
a pseudo-spectral method is used at each time step of the Euler scheme.
This method efficiently computes the quantity
\begin{eqnarray}
\widehat{\chi}_{q_{n}}&=&-\frac{\lambda}{2L} \sum_{m,m'=-N/2}^{N/2} q_m q_{m'}
\widehat{h}_{q_{m}}(t) \widehat{h}_{q_{m'}}(t) \delta_{n,m+m'},
\label{exp:kpzfournonlinear}
\end{eqnarray}
whose real and imaginary part are the non-linear contributions 
of the $N+1$ Langevin equations ~(\ref{exp:kpzfouralpha}) and 
(\ref{exp:kpzfourbeta}). In this procedure, quantity 
$\widehat{\chi}_{q_{n}}$ can be evaluated
without explicitly performing  the double
sum appearing in Eq.~(\ref{exp:kpzfournonlinear}) for each $n$. First,
 the Fourier modes  of the surface derivative  $\partial_{x}
h(x,t)$ are obtained by   a simple  algebraic multiplications $ iq_{n}
\widehat{h}_{q_{n}}$  for $ 0 \le n \le N/2 $. One  then
returns to real space
\begin{equation}
\partial_{x} h(x,t) =  \frac{1}{L} \sum_{n=-N/2}^{N/2} i q_{n}
\widehat{h}_{q_{n}}(t)~ e^{{i}q_n x}.
\label{fourgrad1}
\end{equation}
 to obtain  the gradient at given spatial points.  The computation  of
$\frac{\lambda}{2} (\partial_x h )^2 $  is then straightforward
at these points. One then exploits again a Fourier transform 
to go back to spectral space using the values of the nonlinear terms  
at these collocation points.
In order to prevent the aliasing problem ~\cite{Canuto88},~
\cite{Numericrecipes},
we suitably choose the range of the collocation points and the
number of Fourier modes to be used, in such a way that 
the procedure   provides the exact  $\widehat{\chi}_{q_{n}}$
for $ 0 \le  n \le
N/2 $   of Eq.~(\ref{exp:kpzfournonlinear}). 
The   modes external to the
range $ -N/2 \le n \le N/2$ can be dropped since they do not enter in the 
actual computations.
This well-known technique 
(dealiasing~\cite{Canuto88}), albeit seemingly more complicated,  turns
out to be much more efficient of a brute force computation of 
Eq.~(\ref{exp:kpzfournonlinear}).

\subsection{Comparison with real space discretization}

We now  compare the performance of the  spectral  discretization 
with the one based on the standard Eq.~ (\ref{exp:kpzdiscr1})
and modified Eq.~(\ref{exp:kpzdiscr2}) real space discretization and with
the corresponding analytical results obtained for the continuum KPZ
equation.  
The surface is grown  {\it via} the pseudo-spectral KPZ
equations  with a regularization at scale $a$. 
In these simulations we have always
used an Euler time step in the range $10^{-2}-10^{-3}$ 
which is sufficiently small 
to avoid any numerical
instability up to the sizes and for the $\lambda$ considered in here.

\begin{figure}[tbp]
\centerline{ \epsfxsize=3.5truein \epsfysize=3.5truein
\epsffile{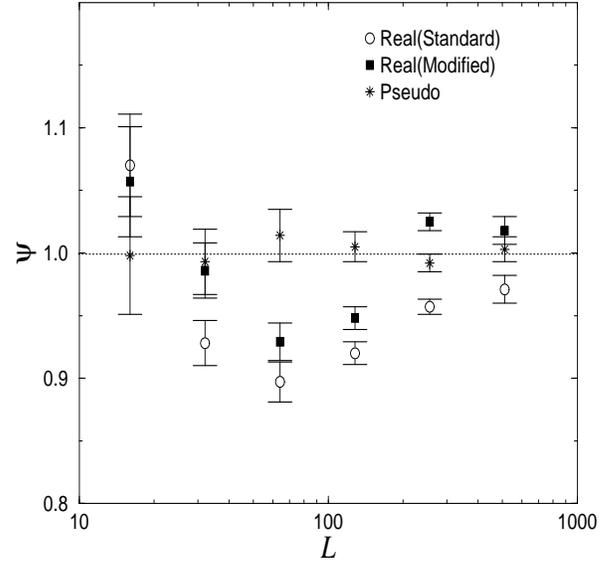} }
\caption{ $\psi(L)=\sqrt{\frac{12\nu}{DL}} W(L)$ 
as a function of the  size $L$
where $W(L)$  stands for the steady state roughness computed using the
standard real space discretization Eq.(\ref{exp:kpzdiscr1}), the
modified real space discretization Eq.(\ref{exp:kpzdiscr2}),  and the
Pseudo-spectral discretization  given in Eq. (\ref{exp:kpzfouralpha})
and Eq.(\ref{exp:kpzfourbeta})  for a one-dimensional KPZ equation.
The dotted line corresponds to the exact continuum value given by
Eq.~(\ref{exp:roughness}) ($\nu/D=1$). Units are arbitrary.}
\label{fig1}
\end{figure}

In order
to compare the performances of the pseudo-spectral method  with respect to
the real space
discretization schemes, we apply a test akin to the one carried of 
in Ref.~\cite{Lam98_2}.
We  generate a steady state KPZ surface in ($1+1$) dimension, with
lattice spacing $a=1$, and parameters $\nu=1$, $\lambda=3$,
$D=1$, $c=0$, by using  (i) the standard real
space discretization~(\ref{exp:kpzdiscr1}), (ii)  the  real space
discretization~(\ref{exp:kpzdiscr2})  introduced in Ref.~\cite{Lam98_2},
(iii) our
pseudo-spectral discretization. 
The   steady state roughness $W(L)$  obtained by
the three methods
for various sizes $L$   is then compared  with the exact value
\begin{eqnarray}
W(L)&=& \sqrt{\frac{D}{12 \nu}} L^{1/2},
\label{exp:roughness}
\end{eqnarray}
of  the continuum KPZ equation~\cite{Krug92}. Fig.~\ref{fig1} shows that,
unlike the standard real space representation~(\ref{exp:kpzdiscr1}) which
underestimates the ratio $\nu/D$ ($=1$ in the present case), both  the
modified real space representation~(\ref{exp:kpzdiscr2})  
and the pseudospectral
representation yield, on average, the correct ratio.
This is no surprise since we  have previously
shown that the pseudo-spectral representation  correctly  accounts for the
steady-state distribution properties, a feature not shared by  
of  the  standard real
space representation ~(\ref{exp:kpzdiscr1})~\cite{Giacometti00}.

\begin{figure}[tbp]
\centerline{ \epsfxsize=3.5truein \epsfysize=3.5truein
\epsffile{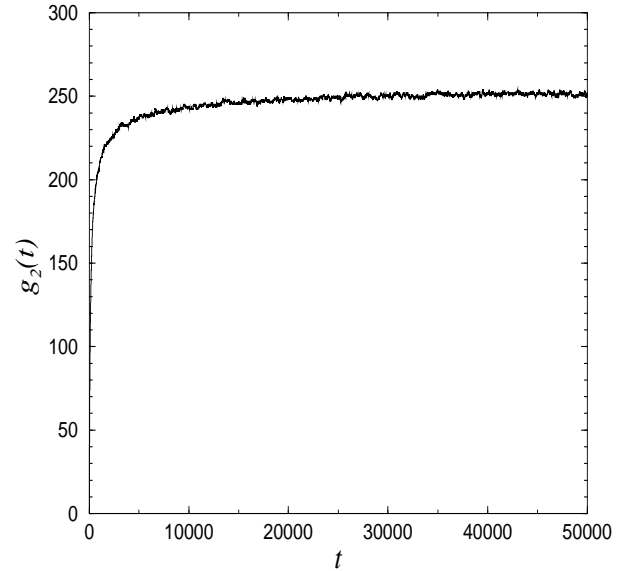} }
\caption{Temporal behavior of $g_2(t)$ for  a KPZ growth on  one
dimensional substrate of size $L=256$.  The temporal axis $t$ is
written in terms of numbers of Euler time steps, here taken
to be $10^{-3}$.}
\label{fig2}
\end{figure}

Figure~\ref{fig2} depicts the behavior of $g_2(t)$ for a surface 
of a size $L=256$, flat at $t=0$, and average  
over ${\cal N}=500$  independent  growths.  The  curve  has a  gradual
increase   (starting   from  zero)   until   it   saturates  after   a
characteristic   time  $t_{\infty}^{\mathbf{c}} \sim 30000 \times 10^{-3}$
This time,   which
depends upon the  size of the system, represents  the cross-over after
which (i) all  length scales have saturated, (ii)  the velocity of the
average height  becomes constant and  (iii) the roughness  levels off.
From renormalization group theory, one expects that 
$t_{\infty}^{\mathbf{c}}\sim L^z$ where the dynamical exponent 
$z$ is equal to $3/2$ for $1+1$
dimension~\cite{Krug97}.

\begin{figure}[tbp]
\centerline{ \epsfxsize=3.5truein \epsfysize=3.5truein
\epsffile{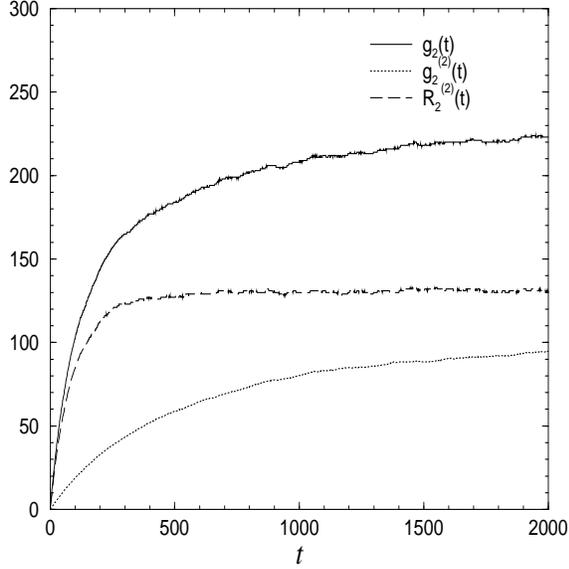} }
\caption{Temporal behavior of functions $g_2(t)$,   $g_{2}^{(2)}(t)$
and $R_{2}^{(2)}(t)$  (see text) for a KPZ growth on  one dimensional
substrate of size $L=256$.  The temporal axis $t$ is written in terms
of numbers of Euler time steps $10^{-3}$.}
\label{fig3}
\end{figure}

In the spirit of renormalization group theory, let us separate the dynamics
of modes of  wavelength shorter than $a_s=ba$ with  $b=2^l$ with modes
of wavelength longer than $a_s$.  This can be done using the following
decomposition of $g_{2p}(t)$ ( $N_s=N/b$)
\begin{eqnarray}
g_{2p}(t)&=& g_{2p}^{(b)}(t) + R_{2p}^{(b)}(t),
\end{eqnarray}
with
\begin{eqnarray}
g_{2p}^{(b)}(t)&=& \frac{1}{L}  \sum_{n=-N_s/2}^{N_s/2} q_n^{2p} \left
\langle |\widehat{h}_{q_n}(t)|^2 \right \rangle.
\label{g2sat}
\end{eqnarray}
representing the ``slow'' part, and
\begin{eqnarray}
R_{2p}^{(b)}(t)&\equiv&    \frac{1}{L}   \sum_{n=-{N}/2}^{-N_s/2   -1}
q_n^{2p}  \left  \langle  |\widehat{h}_{q_n}(t)|^2  \right  \rangle \\ \nonumber
&+&
\frac{1}{L}\sum_{n={N_s}/2    +1}^{N/2}    q_n^{2p}   \left    \langle
|\widehat{h}_{q_n}(t)|^2 \right \rangle ,
\end{eqnarray}
indicating the ``fast'' part.
In Fig.~\ref{fig3},  the behavior  of $g_2(t)$ ,  $g_{2}^{(b)}(t)$ and
$R_{2}^{(b)}(t)$  are   plotted  for   a  scaling  $b=2$.    Before  a
characteristic time  $t_1^{\mathbf{c}} $ (  $t_1^{\mathbf{c}} \sim 350
\times 10^{-3}$  for the parameters chosen in  the pseudo-spectral KPZ
equations), shortwave  modes $N_s/2 +1  \le |n| \le N/2$  contained in
$R_{2}^{(2)}(t)$ are  evolving much faster than longwave  modes $0 \le
|n| \le N_s/2$.   Similar features occur for higher  values of $b=2^l$
thus     defining    a     sequence     of    characteristic     times
$t_1^{\mathbf{c}}<t_2^{\mathbf{c}}<t_3^{\mathbf{c}}<\ldots$.        For
instance it is found that $t_2^c \sim 1300 \times 10^{-3}$ and
$t_3^c \sim 3000 \times 10^{-3}$.  
The  above remarks are  clearly important when  defining the
dynamics  of the  coarse-grained surface  obtained by  eliminating the
modes of wavelength shorter  than scale $a_s=ba=2^la$. This surface is
characterized  by  the  same  average  height $g_0^{(b)}  =  g_0$  
for any $b$, and
$g_{2}^{(b)}(t)$ playing the role of  $g_2(t)$.     By     rewriting
Eq.~(\ref{ensaverage6}) as
\begin{eqnarray}
\frac{d g_0^{(b)}(t)}{dt} &=& c \biggl[ 1+ \frac{\lambda}{2L}R_{2}^{(b)}
\biggr]L + \frac{\lambda}{2}g_2^{(b)}(t),
\label{renorm1}
\end{eqnarray}
it is clear that the coarse-grained surface with $N_s=N/b$
modes  will satisfy a KPZ  equation with  a renormalized $c_s$ and 
identical  $\lambda_s=\lambda$ only when $R_{2}^{(b)}$ has 
reached saturation.
On starting from a flat interface, this
happens  whenever the Fourier modes   $N_s/2 <n< N/2 $ and 
$ - N/2 < n < -N_s/2$ are  thermalized, i.e. for   $t >t_l^c$.
 Similarly, upon summing  Eq.~(\ref{ensaverage3}) over  the slow modes
 $n=1,\ldots,N_s/2$, one finds that,
\begin{eqnarray}
\frac{d g_2^{(b)}(t)}{dt} &=& -2 \nu  g_4^{(b)}(t) - \lambda
\sum_{n=-N_s/2}^{N_s/2} q_n^2 {\cal V}_{q_n} + 2D Q^{(b)}_2,
\label{renorm2}
\end{eqnarray}
where
\begin{eqnarray}
Q^{(b)}_{2} &=&   \sum_{n=-N_s/2}^{N_s/2} q_n^2.
\end{eqnarray}
For $t\ge t_1^{\mathbf{c}}$, it is assumed that 
dynamics of fast modes is slaved to the one of slow modes, in such a way that
the term in  $\lambda$  can be written as   
\begin{eqnarray}
\lambda \sum_{n=-N_s/2}^{N_s/2} q_n^2 {\cal V}_{q_n} =  2 \Delta \nu
g_4^{(b)}(t) - 2 \Delta D Q^{(b)}_2,
\end{eqnarray}
thus satisfying a coarse-grained KPZ equation with  renormalized 
parameters $\nu_s= \nu +\Delta \nu $ and $D_s= D
+ \Delta D$.
For the EW universality class, it is easy to show that,
using Eq. (\ref{renorm1}) and Eq. (\ref{renorm2}), all parameters
do not renormalize, as expected from renormalization group theory
\cite{Marsili96}.

Our purpose here is to  compute the KPZ renormalized
parameters directly from the
experimental observations of the surface growth. According to the previous
discussion, we should begin collecting the data after a time
$t \ge t_l^{\mathbf{c}}$ (starting from an initially flat surface)
in order to best characterize the dynamics of a surface
coarse-grained by a factor
$b=2^l$. Moreover the most prominent evolution  of this surface
occurs during the time interval when the length scales 
between $2 a_s$ and $a_s= b a$
are not thermalized  i.e. during the  interval
$[t_{l}^{\mathbf{c}},t_{l+1}^{\mathbf{c}}]$ which  is then the
optimal period to characterize its  dynamics.

\section{The reconstruction procedure based on the  spectral approximation.}
\label{sec:ident}
Based upon the above spectral  approximation, we now introduce a method   
to identify, at a given length scale, an optimal set of KPZ
effective coupling parameters $c$, $\nu$, $\lambda$ and $D$,   from
"experimental" snapshots of  interface profiles.  The  "experimental"
data may be generated by numerical simulations of the KPZ
equation itself~(see tests in~section~\ref{KPZpseudoreconstr}), by
numerical microscopic models emulating surface
processes~(Section~\ref{sec:Micro}) or associated with 
real  interface growth. For the
sake of  simplicity,  measurements are always assumed free of
observational noise.  This approach thus constitutes a typical inverse
problem for  an infinite (or finite, but large, in the discretized
approximation) dimensional system with a finite number of parameters
to identify. Although in this work we focus  on the KPZ universality class,
our method has a more general validity, and may be extended, with slight
changes, to other universality classes.

A first  reconstruction of KPZ dynamics was attempted  by Lam and
Sanders~\cite{Lam93}. These authors used
equation~(\ref{exp:kpzdiscr1}) and worked in real space using 
experimental  heights $h_i^{obs}(t)$ measured at $N$ points $x_j= j a$
with  $j=1,...N$.  For the reconstruction, they performed
a least-squares calculation 
directly on the Langevin equations  to   compute the
parameters $c$, $\nu $, $\lambda$. The  noise term  $D$ was eventually
obtained  as a by product of the previous calculation.  
In Ref.~\cite{Giacometti00}, the difficulties associated with this
approach have already been discussed.
In the present  work,  the analysis is  based  on the philosophy
explained in   previous Sections and
equations~(\ref{ensaverage4}), respectively (\ref{ensaverage6}), 
are  used   to
identify through a least-squares procedure  the  coefficients $\nu $,
$D$, respectively $c$, $\lambda$.
Besides the fundamental theoretical features discussed in Sec.~\ref{sec:KPZ},
several reasons may be invoked in favor of  our  approach.
First, this method  is  not directly  based on the primitive
stochastic equations but  uses the deterministic equations introduced
in Section~\ref{sec:correlations} which govern the  ensemble average
of correlation functions.  
Standard identification algorithms (e.g. the least-squares method)
which are well suited and have been widely tested for deterministic equations,
are then expected to be more reliable under such conditions.
Second, as functions $g_{2p}(t)$ are already 
averaged quantities, a smaller number of realizations  
is expected to be required due to the self-averaging property.
Finally, dynamical noise is  directly introduced in our reconstruction
algorithm, unlike in the one of Ref.~\cite{Lam93}. 
This  seems a natural requirement since noise is an intrinsic
parameter of the interface evolution.

In order to get  ensemble averages of  spatial correlations,  ${\cal
N}$ growths starting from the same surface, e.g. a flat surface, have
been  carried out. This provides ${\cal N}$ distinct realizations of
the same  stochastic process. For a  given realization,  the
experimental surface is  observed at time $ \left( t_k =k\Delta t,
k=1,2,..,pM \right)$ where $\Delta t$ is the measurement sampling
time. This  procedure  may  be easily performed in a real experiment
and leads to  quantities  $g_0(t)$, $g_2(t)$ and $g_4(t) $   which are
linked, at the scale considered,  to the average height, the average
of the square of the  first or second  surface derivatives of the
smoothed surface.

Let us now  integrate equation~(\ref{ensaverage4}) during $p$ sampling
times $\Delta t$~:
\begin{eqnarray}
{ \frac { g_2(T_{k+1})-g_2(T_{k})} {T_{k+1} - T_{k}} } &=&  -2 \nu
{\frac {1} {T_{k+1} - T_{k}} }  \int_{T_{k}}^{T_{k+1}} g_4(t) dt +
2D Q_{2},
\label{rec1}
\end{eqnarray}
where time  $T_{k} =k  p \Delta t$, $(k=1,\ldots,M)$. Similarly
from equation~(\ref{ensaverage6}) one gets
\begin{eqnarray}
{ \frac { g_0(T_{k+1})-g_0(T_{k})} {T_{k+1} - T_{k}} } &=& cL+
\frac{\lambda}{2} \frac {1} {T_{k+1} - T_{k}}   \int_{T_{k}}^{T_{k+1}}
g_2(t) dt.
\label{rec2}
\end{eqnarray}
If  $\Delta t$ is  smaller    than  the characteristic time of the
dynamics,   one may  approximate  the   time integral in Eqs.~(\ref{rec1})
and (\ref{rec2}),  as an  average  over the $p$ intermediate
sampling times, thereby obtaining $M-1$ linear constraints on the
parameters $\nu $, $D$ and $c$, $\lambda$.  A simple least-squares
calculation then determines  $\nu $, $D$ from Eq.~(\ref{rec1})
and $c$, $\lambda$ from Eq.~(\ref{rec2}).

\section{Results on the KPZ reconstruction}
\label{KPZpseudoreconstr}

In order to test our reconstruction method,  we use the above
procedure on rough surfaces  generated through a discretized  KPZ
equation with known coupling parameters ($\nu=1$,$\lambda=3$,$D=1$, $c=0$). 
Note that, in order to
compute the ensemble average values  $g_0(t)$, $g_2(t)$ and $g_4(t) $,
we use ${\cal N}=500$ samples to obtain a convergent value.  This is important
since statistical fluctuations may be large enough to induce  a
measurement error  exceeding the non-linear contributions, and under such
circumstances the identification would fail.  Furthermore each
reconstruction is repeated for a small number of independent
configurations (typically $5$) in order to get an estimate of the
error bars associated with the reconstructed parameters.

In the absence of  coarse-graining,  data are produced   by the
discretized  equations~(\ref{exp:kpzfourf0})-(\ref{exp:kpzfouralpha})-
(\ref{exp:kpzfourbeta}) where the minimum length scale $a  = L/N$  is
introduced by the spectral approximation. Since  equations used in the
reconstruction procedure are  exactly identical to the ones
generating the time series, this provides a first stringent test on
the validity of our  method.  For the  original data,   the most
efficient interval choice to perform the identification   is  
expected to be
$[0,t_{1}^{\mathbf{c}}]$.  Indeed, after $t_1^{\mathbf{c}}$,  all
length scales between $a$ and $2a$ are thermalized   and the
contribution to the variation of  steady
velocity, stemming from the  non-linear term, 
becomes  less and less  important (see
equation~(\ref{renorm1})).  In this instance, 
the inversion technique may run into
difficulties to compute $c$ and  $\lambda$ since   statistical
fluctuations in the computation of ensemble averages should not be
greater  than the value of this nonlinear contributions, as mentioned
above.

\begin{figure}[tbp]
\centerline{ \epsfxsize=3.5truein \epsfysize=3.5truein
\epsffile{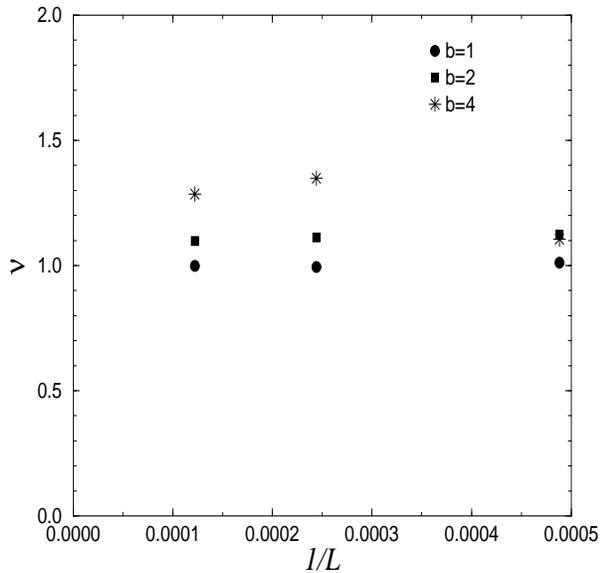} }
\caption{The coupling parameter $\nu_s$ for various coarse-graining
levels $b=2^l$ and for increasing   lattice sizes
$2048$, $4096$ and $8192$ in the case of a numerically generated KPZ
equation. The input  value is $\nu=1$. Error bars are of the
order of the symbol sizes.}
\label{fig4}
\end{figure}

\begin{figure}[tbp]
\centerline{ \epsfxsize=3.5truein \epsfysize=3.5truein
\epsffile{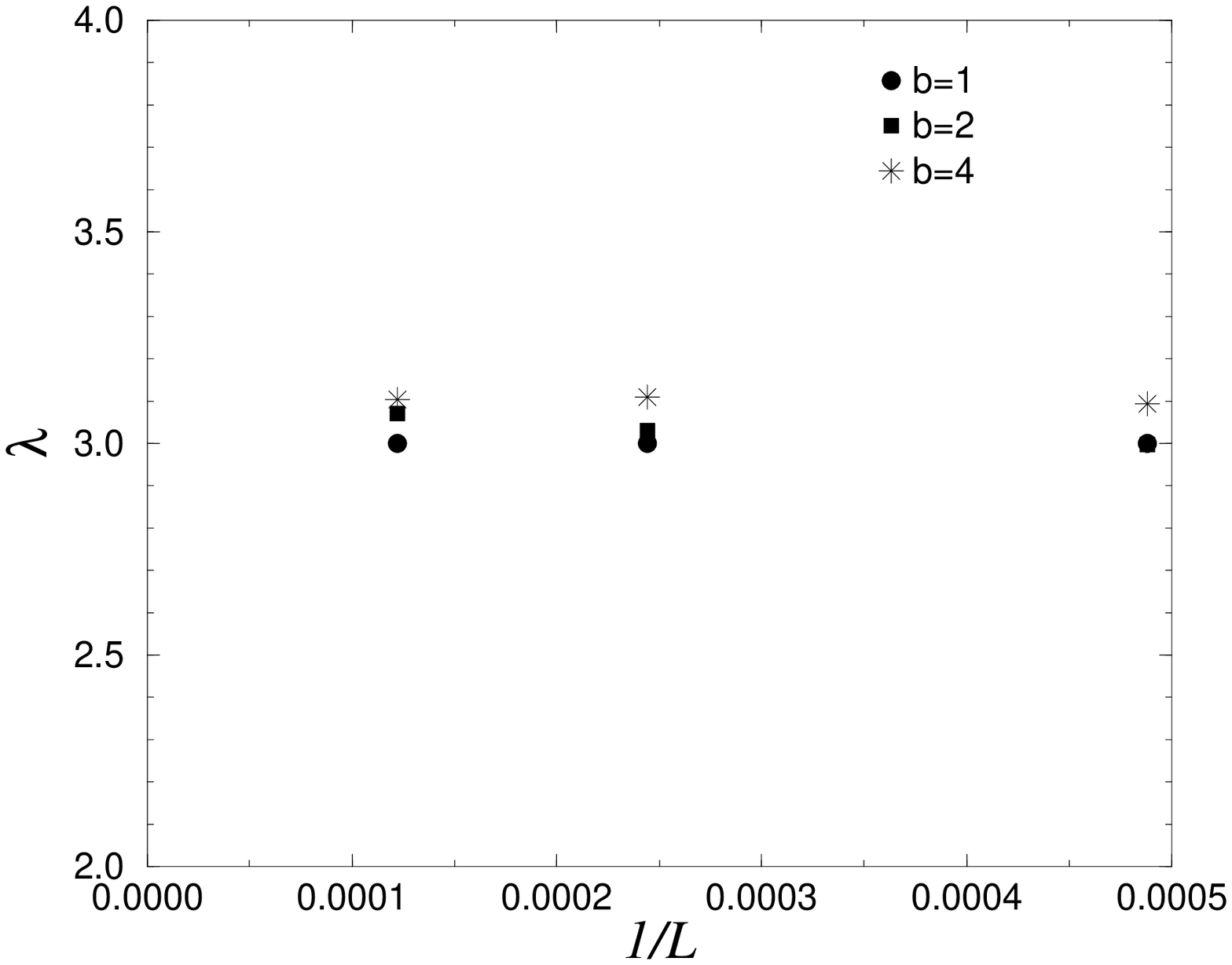} }
\caption{The coupling parameter $\lambda_s$ for various $b$
and increasing   lattice
sizes $2048$, $4096$ and $8192$ in the case of a numerically generated
KPZ equation. The input value is $\lambda=3$.}
\label{fig5}
\end{figure}

\begin{figure}[tbp]
\centerline{ \epsfxsize=3.5truein \epsfysize=3.5truein
\epsffile{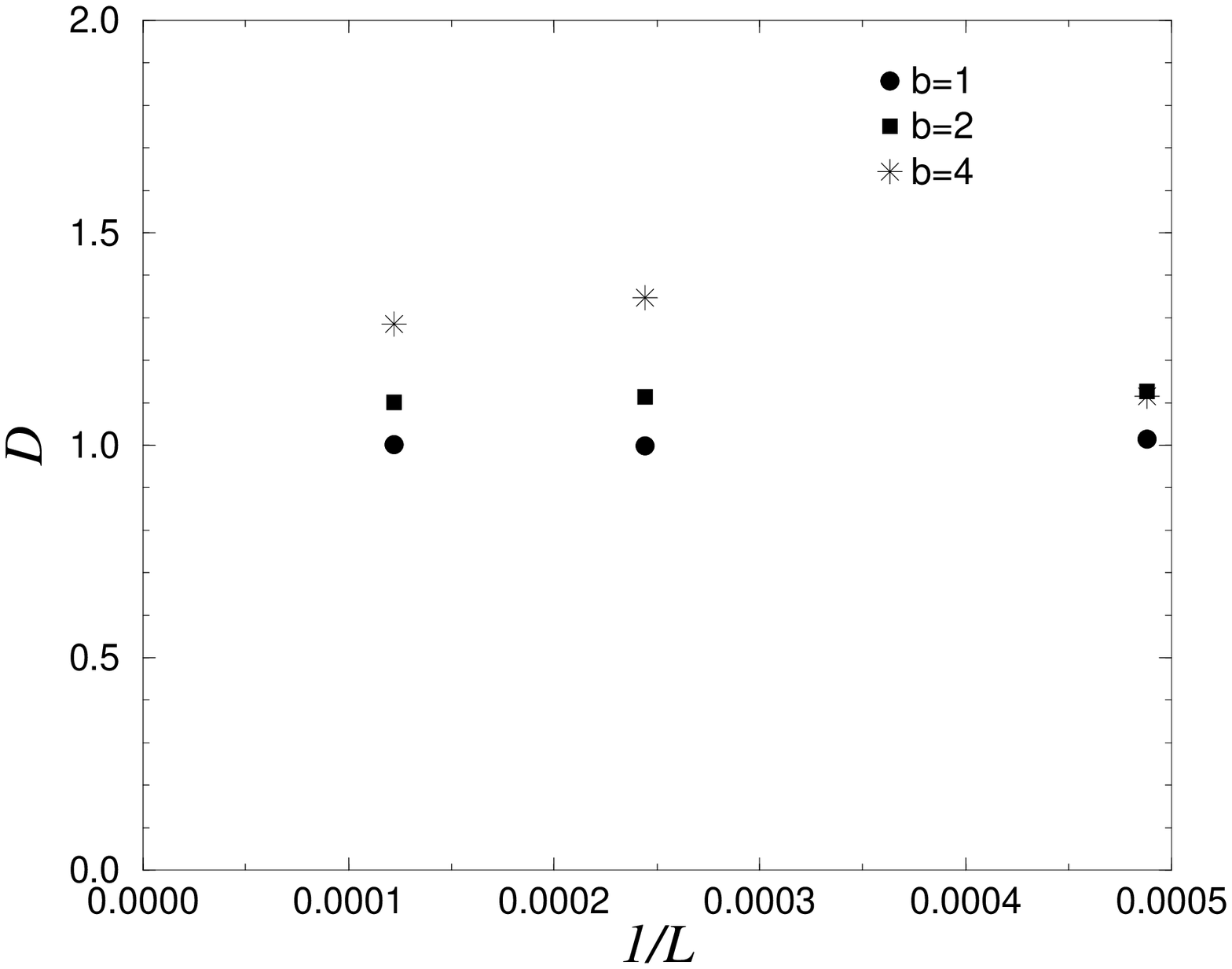} }
\caption{The coupling parameter $D_s$ for various $b$ and
for increasing   lattice sizes
$2048$, $4096$ and $8192$ in the case of a numerically generated KPZ
equation. The input value is $D=1$.}
\label{fig6}
\end{figure}

Such a  procedure may be iterated for coarse-grained surfaces at
$a_s=ba$ ($b=2^l=2,4,\ldots$) which  are still assumed to be governed
by KPZ-like equations.  The reconstruction of the renormalized
equation  should then be performed with data taken after
$t_l^{\mathbf{c}}$ since the renormalized equation is not expected to
be valid at earlier times. Once again, because of statistical
fluctuations, we expect the most efficient interval choice for the
identification at length scale $a_s$  to be  $[
t_{l}^{\mathbf{c}},t_{l+1}^{\mathbf{c}}]$.           Clearly the
reconstruction  becomes more and more  difficult to implement as
$b$ increases, since time intervals
$[t_l^{\mathbf{c}},t_{l+1}^{\mathbf{c}}]$  and required statistics
will correspondingly increase. 
Fig.\ref{fig4}, \ref{fig5}, \ref{fig6} depict the
results for the  three parameters $\nu_s$, $\lambda_s$ and $D_s$  as a
function  of the scaling ratio $b$ for various lattice  sizes (error bars
are of the same order of magnitude of the symbol sizes and therefore
not shown). The
optimal values is then expected to correspond to the $L \to \infty$
limit.   For the scale $a$, one obtains      the correct values which
shows the capability of the method to identify the correct parameter.
For the coarse-grain case the extrapolated values of $\lambda$,  
as well as the ratio $\nu/D$,  appear to be independent of $b$
(as they should) 
while    parameters $\nu$ and $D$ are
renormalized  \cite{Barabasi95}.

\section{Results on growth models}
\label{sec:Micro}

The ultimate goal for our method is to be applied to {\it real   
experimental}  surfaces.   In this
section, as an intermediate step, 
the reconstruction  technique is tested  on  data
produced through numerical   microscopics models.  Specifically we
consider  two typical models: (i) a random deposition model with surface
diffusion (RDSD) which is  described by the EW linear continuum theory
($\lambda =0$ ); (ii) a particular solid-on-solid model called
single-step1 (SS1) which is expected to belong to the KPZ universality
class.  This latter model can be mapped onto
an Ising model \cite{Krug92} thus providing  an analytical
value of $\lambda$ and hence a further stringent test for our method.

A microscopic growth model  is  composed of  three main
ingredients. (L1) A probabilistic law, independent of the  surface
dynamics  itself, which describes    the flux of particles directed
towards  the surface; (L2) a -- deterministic or probabilistic -- rule
which  determines whether a given  particle directed towards
the specific site $s$ is effectively deposited ($s$ is an active site)
or simply discarded($s$ is not   active); (L3) a  prescription yielding the
displacement of the particle or the rearrangement of the surface,
after that the deposition has become effective.

Law (L1), characterizing the particle flux, defines the mean
time or the time scale necessary to  a particle to  fall  towards --
but  not necessarily deposited on -- the surface. In principle, 
this law may be
given by a   probability varying with site locations and may also be of
intermittent nature.  In this work, however, we consider the simplest  case of
uniform flux both in space and time.  During  each time interval
$\delta t$,  which defines the time scale of the process,  a unique
particle   falls  on   the surface with probability one and selects a
given  site $s$ with equiprobability $1/N$.

In order to compute the parameters, we use as space unit the unit cell
of the microscopic model  $a=1$ ($L=N$ where $N$ is the number of
sites).  The unit of time is defined in such a way that 
$\delta t =1/N$. In the following, we use sizes up to
$L=8192$ with ${\cal N}=50$ independent growths.

\subsection{The RDSD  model.}

In the RDSD,   a particle directed towards the specific site $s$ is
always deposited (L2).   This means that, one layer is  deposited 
on the surface during a unit of time.
Law (L3) can be phrased as follows. Upon reaching the surface,  the
particle falling towards a specific site $s$, compares the
heights of the nearby neighbors and sticks to the one of lowest
height unless the original site is a local minimum (in that case it
does not move).

We have generated a RDSD starting from a flat
surface and performed coarse graining up to $b=16$.  The  sampling
measurement time   $\Delta t$ is taken to be larger then the time
unit. We have
consistently found $\lambda \sim 0$ and $c \sim 1$, as expected. 
Moreover we find a value for $\nu$ and
$D$ which, as $b$ increases, slowly converge to 
$\nu \sim 0.8$ and $D \sim 0.5$ (we recall that $c=1$ and $D=0.5$ for a simple
random deposition model with our time units).

\subsection{The SS1 model.}

For the SS1 model,  an ``active site''  is defined~\cite{Meakin86} as
a site  which is a  local minimum for nearby sites, i.e. such that
$h_i < h_{i-1}$ and $h_i < h_{i+1}$ (L2) .  The (L3) rule
for the SS1 model imposes that   two particles are deposited
on the active site. Our algorithm has been devised to cover non-steady-state
conditions. Since this situation is hardly discussed in the literature
\cite{Barabasi95},\cite{Meakin86},
in Appendix~\ref{sec:appB} the way of including the time dependence in this
model in an efficient way, is reported.
The  sampling measurement
time   $\Delta t$ is taken to be the unit  of time.
Note that, unlike the
RDSD model,  the time unit does not correspond to the  effective 
deposition  time
of one layer.

\begin{figure}[tbp]
\centerline{ \epsfxsize=3.5truein \epsfysize=3.5truein
\epsffile{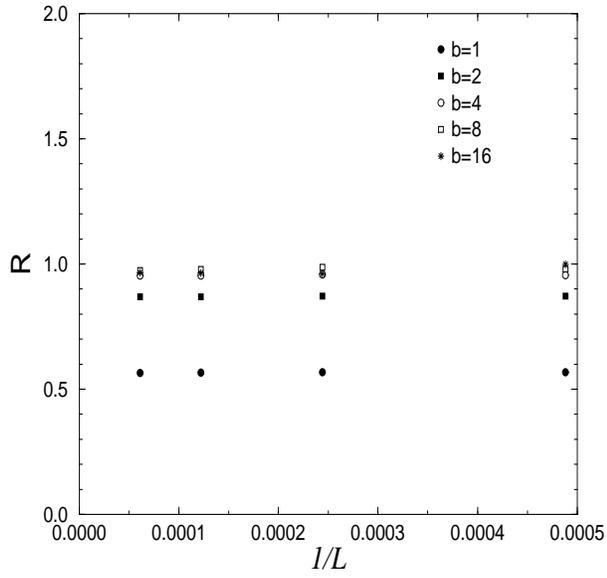} }
\caption{The ratio $R \equiv \nu_s /D_s $ for the SS1 growth model.}
\label{fig7}
\end{figure}

\begin{figure}[tbp]
\centerline{ \epsfxsize=3.5truein \epsfysize=3.5truein
\epsffile{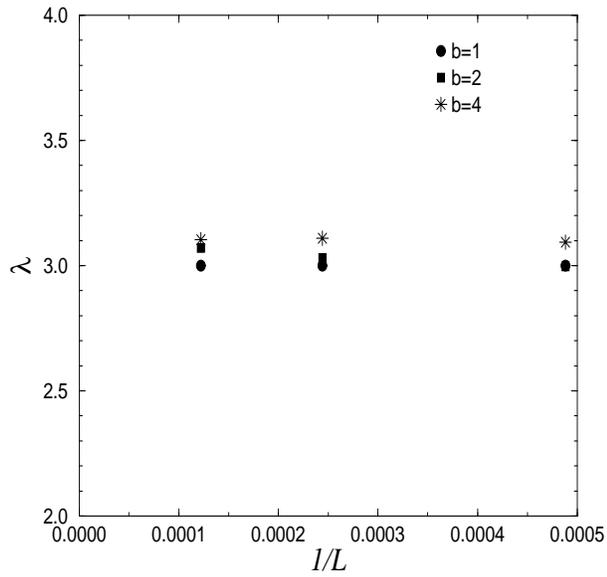} }
\caption{The coupling parameter $\lambda_s$ for the SS1 growth model.}
\label{fig8}
\end{figure}

We have used a ``teeth-like'' initial surface with  odd and even sites having
height 0 and 1 respectively (hence there are
$N/2$ active site at this stage) \cite{Meakin86}. 
Reported in Fig.~\ref{fig7} is the
result for the ratio  $\nu_s/D_s$ for various scaling factors $b$. 
As $b$ increases,   
$\nu_s/D_s$ tends to a constant characteristic of the KPZ phenomenology
for $b \ge 4$. 
This is confirmed by Fig.~\ref{fig8} which depicts the results for
$\lambda_s$ for $b \ge 4$, displaying a tendency for $\lambda_s$ towards $-2$.

A word of caution is in order here. As we discussed, the value
of $\lambda$ is exactly known
through a mapping onto an Ising model \cite{Krug92}. The predicted value
is $\lambda=-1$ when computed with a time unit yielding $c=1$
at stationarity. This result  is consistent 
with ours since it  corresponds to a time unit which 
is half of the one we have exploited ($c_s \sim 0.5$).

\section{Conclusion.}
\label{conclusion}
In this work we have proposed a new method to extract the coupling
parameters of the KPZ equation from experimental snapshots of 
successive interface profiles.
This method hinges on two main ingredients. First 
a pseudospectral scheme is used to simulate the KPZ equation, and this
scheme  
can be reckoned as an improved discretization with respect
to the standard real space finite difference ones. As a matter of
fact, it preserves both the correct steady state distribution
and the coarse-graining properties of the continuum corresponding
equation. Second, our reconstruction algorithm is based on
the time evolution of correlation functions. These functions 
do satisfy a deterministic evolution equation entitling 
the use of standard least-square procedure to identify the
coupling parameters. This second ingredient parallels the analogous one
introduced in Ref.\cite{Giacometti00} which was however based on a real
space representation.

We have first tested the overall procedure on numerically generated
KPZ profiles with known coupling parameters. Our scheme is capable 
to reproduce 
not only the correct parameters in the absence of coarse-graining but
unlike the previous 
attempt \cite{Giacometti00}, it  also
provides consistent and robust results for coarse-grained surfaces.

Next we have applied our algorithm to microscopic models which more
closely mimic experimental situations. In such a case, a smoothing
procedure is unavoidable in order to describe the surface in terms
of a continuum evolution equation, and it is then a vital requirement
to use an efficient and reliable method under such conditions.
Again we were able to reproduce few known analytical results
for these microscopic growth models. Furthermore, some additional
estimates of other parameters have also been given. 

We remark that our method is of general applicability. For instance, 
an extension to two-dimensional surfaces is not expected to present
major difficulties. Similarly it could be applied 
to determine stochastic equations  emulating  coarse
grained equations of the Kuramoto-Shivashinski type where
the universality class is still an open question \cite{Boghosian99}. 

\acknowledgments

Funding for this work was provided by a joint CNR-CNRS exchange
program (number 5274), MURST and INFM.  It is our pleasure 
to thank Rodolfo Cuerno, Matteo Marsili and Lorenzo Giada
 for enlighting discussions.

\appendix

\section{Proof that $K(t)=0$}
\label{sec:appA}

We  prove here that the quantity $K(t)$ appearing in 
Eq.~(\ref{ensaverage4first}) is
actually zero. The proof is patterned after a similar proof used to
show that the steady state probability (\ref{exp:kpzsteadyfou}) is
independent of $\lambda$ for a ($1+1$) KPZ equation.  Indeed by a
reflection $q_n \to - q_n$ we obtain
\begin{eqnarray}
K(t)&=& \frac{1}{L^2} \sum_{n,m,m'=-N/2}^{N/2} q_n^2 q_m q_{m'} \\ \nonumber 
& &
{\cal R}
\left \langle \Bigl( \widehat{h}_{q_n}(t) \widehat{h}_{q_m}(t)
\widehat{h}_{q_{m'}(t)} \Bigr) \right \rangle \delta_{0,n+m+m'}.
\end{eqnarray}
The above quantity is invariant under a cyclic permutation of the
indices $n,m,m'$. Therefore it can be  rewritten  as:
\begin{eqnarray}
K(t)&=& \frac{1}{L^2} \sum_{n,m,m'=-N/2}^{N/2} \frac{1}{3}q_n q_m
q_{m'} \bigl[ q_n+ q_m+ q_{m'} \bigr] {\cal R} \left \langle  \Bigl(
\widehat{h}_{q_n}(t) \widehat{h}_{q_m}(t) \widehat{h}_{q_{m'}(t)}
\Bigr) \right \rangle \delta_{0,n+m+m'},
\end{eqnarray}
which obviously vanishes.

\section{Time evolution of the  SS1 deposition model: an efficient algorithm}
\label{sec:appB}

A naive simulation of the SS1 model would proceed as follows. During
each time interval $\delta t$,  a unique
particle is dropped on the surface, and one checks whether it is falling on an
active or inactive site according to (L2).  If deposition is attempted
on an active site,  according to law (L3) for SS1, 
time is incremented and the particle is deposited.   
On the contrary,  the particle directed towards  an
inactive site is discarded  but the time  is
nonetheless incremented.  This procedure   is  however very time
consuming. Efficient algorithms,  generating  equilibrium
surfaces  in a  fast way,  simply  consider  active sites 
and do not take time into account.
Such  algorithms  cannot be used here since (i) we have not
reached equilibrium and (ii) we  need to quantify the interface
effective time evolution.  Therefore we next implement an improved
algorithm providing the time evolution as well.

At time $t= k \delta t$, let  us assume we know the number 
$ N_a(t)$ of active sites of the surface and  their respective positions. We
then select, with equal probability, one of  such active sites, 
we deposit a particle on it and then perform the
re-ordering of the surface and the updating of  the active sites. 
The important question is how long we have to wait
to see the deposition event to occur. 
The probability of deposition between $t$ and  $t+ \delta t$
is  $\alpha \equiv N_a(t) /N$ and the probability 
that $k=t/\delta t$ time steps elapse before a deposition occurs is 
 $\alpha [1-\alpha]^{k-1}$. This law of
probability of time intervals  is  numerically generated as follows.
Intervals $[\theta_0,  \theta_1]$,
$[\theta_1,\theta_2]$,...$[\theta_k,\theta_{k+1}]$  are defined  in
$[0,1]$ where $\theta_0=0$,  $\theta_1=\alpha$ and
$\theta_{k+1}=\theta_{k}+ \alpha [1-\alpha]^{k}$, i.e.  $ \theta_{k} =
1- [1-\alpha]^k  $.  Assume that  a number $\beta$  is chosen with an
equiprobability in the interval $[0,1]$. If  it lies in the interval
$[\theta_k,\theta_{k+1}]$, then the  waiting time   is given by
$(k+1)\delta t$.  


\begin{figure}[tbp]
\centerline{ \epsfxsize=3.5truein \epsfysize=3.5truein
\epsffile{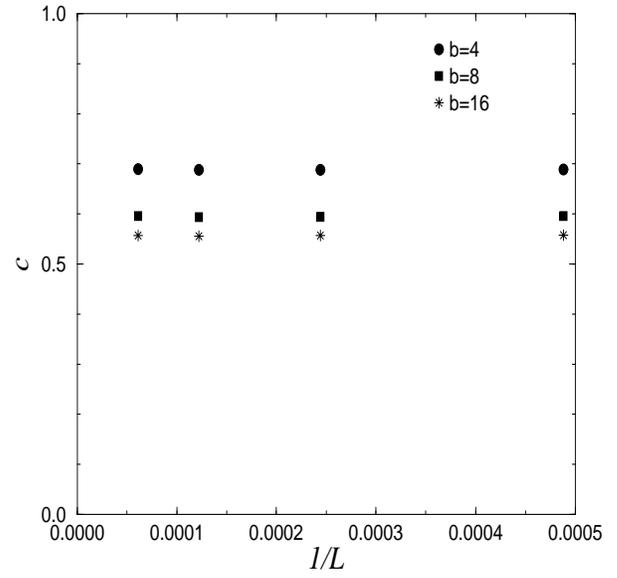} }
\caption{The coupling parameter $c_s$ for the SS1 growth model.}
\label{fig9}
\end{figure}

\end{document}